\documentclass{myws-p10x7}

\begin{document}

\title{Measurements of the bottom quark mass}

\author{G. Dissertori}

\address{EP Division, CERN, CH-1211 Geneva 23, Switzerland
         \\E-mail: Guenther.Dissertori@cern.ch}

\twocolumn[\maketitle\abstract{
I will review new measurements of the $b$ quark mass, presented at this
conference by ALEPH and DELPHI. A large set of observables has been used 
and detailed studies on jet algorithms have been performed. 
These measurements at the Z peak are consistent with the results obtained
at the $\Upsilon$ scale when assuming the running of the $b$ quark mass
as predicted by perturbative QCD.}]


\section{Introduction}

\mbox{}
\vspace{-0.5cm}

\noindent The $b$ quark mass is one of the fundamental parameters of the QCD Lagrangian.
However, due to confinement, quarks do not appear as asymptotically
free particles and therefore the definition of their mass is ambiguous. 
In the framework of perturbative QCD,
quark masses can either be defined as the position of the 
pole of the quark propagator, or they can be interpreted as effective
coupling constants in the Lagrangian. In the former definition 
the mass is called ``pole mass'' and does not depend on an energy scale; 
in the latter the mass is called ``running mass'', since it is a function
of the renormalization scale.


The effects of the $b$ mass become very small with increasing 
energy for inclusive observables  such as the total cross section,
since they are proportional to 
$m_b^2/M_Z^2$ ($\cal O$ (0.1\%)). However, for semi-inclusive quantities
such as jet rates, the effects are enhanced, up to a 
few percent. Such quantities are sensitive to the amount of gluon radiation,
which is suppressed in the case of massive quarks. 

At this conference new measurements have been presented by ALEPH\cite{ALEPH_paper} 
and by DELPHI\cite{DELPHI_paper}.
Previously measurements 
had been published by DELPHI \cite{DELPHI_orig}
and by Brandenburg {\it et al.} \cite{SLD} who had analysed SLD data.



\section{Analysis Method}

The method for extracting the $b$ quark mass is based on the measurement 
of the ratio $R_{b/uds} = O_b/O_{uds}$ of an infrared safe observable $O$ 
computed for  $b$ and $uds$ induced events and assuming $\alpha_s$ 
universality.

This ratio is either directly obtained by tagging $b$ and 
$uds$ induced  events (DELPHI), or by measuring first the ratio
$R_{b/inc} = O_b/O_{inc}$ ($inc$=all flavours inclusive) and then
inferring from that the ratio $R_{b/uds}$, using the precise 
knowledge of the partial widths of the Z to $b$ and $c$ quarks (ALEPH).
The tag of $b$ ($uds$) events is  mainly based on lifetime information.
The ratio is then corrected for hadronization, detector and tagging biases.


The $b$ quark mass is extracted by comparing the corrected measured ratio $R_{b/uds}^P$ 
with the predictions for the observable under study.
In general the NLO prediction for $R_{b/uds}^P$ as a function of
the $b$ quark mass is of the form
\begin{equation}
R_{b/uds}^P = 1+\frac{m_b^2}{M_Z^2}\left[ b_0(m_b) + \frac{\alpha_s}{2\pi}b_1(m_b)\right]
\end{equation}
where the coefficient functions $b_0$ and $b_1$ are obtained from the
integration of the massive and massless matrix elements in terms of the
pole or running mass.

Systematic uncertainties arise from imperfections of the detector modelling,
from the limited knowledge of the hadronization corrections and B decays, which
are particularly important for this analysis, and from theoretical ambiguities
because of the renormalization scale and the quark mass scheme employed
to compute $b_0$ and $b_1$.


\section{Results from ALEPH }

ALEPH has studied the first and second moments of a large set of 
event shape variables such as thrust, jet broadenings, or the
transition resolution value $y_3$ for going from three to two 
jets when applying the Durham \cite{Durham} algorithm. Furthermore,
they have used the ratio of three-jet rates with $y_{cut} = 0.02$. 
Then the set of 
observables is reduced by requiring that the NLO contributions 
to the perturbative prediction be clearly smaller then the LO terms,
and that the hadronization corrections do not exceed
the measured mass effect in size. These requirements leave only
the three-jet rate and the first moment of $y_3$ as observables.
The latter turns out  to give the smallest total uncertainties, 
and the result for the running mass in the $\overline{\mathrm{MS}}$
scheme is
$
 m_b(M_Z) = (3.27\pm0.22_{stat}\pm0.22_{syst}
    \pm0.38_{had}\pm0.16_{theo})\;
            \mathrm{GeV}/c^2.
$


\section{Results from DELPHI}

They have updated their previous analysis \cite{DELPHI_orig}
by including data from 1995, improving the $b$-tag algorithm,
and by studying also the Cambridge jet clustering algorithm \cite{Cambridge}
for the ratio of three-jet rates. It turns out that with this
algorithm the perturbative expansion for $R_{b/uds}^P$ 
converges more rapidly in the running mass scheme than in the
pole mass scheme, and the theoretical uncertainties are smaller
than with the Durham algorithm. However, 
the measurement based on the Cambridge algorithm
still suffers from rather large hadronization uncertainties.
The result is $m_b(M_Z) = (2.61\pm0.18_{stat}\pm0.18_{syst}
    \pm0.47_{had}\pm0.07_{theo})
            \mathrm{GeV}/c^2.
$


\begin{figure}
\epsfxsize190pt
\epsfysize210pt
\figurebox{170pt}{8cm}{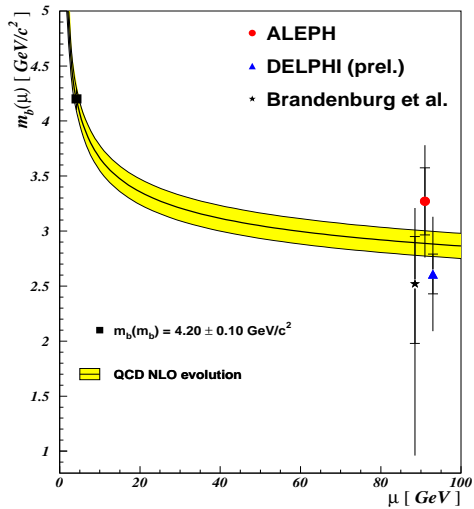}
\caption{The running of the $b$ quark mass}
\label{fig1}
\end{figure}

\section{Conclusions}

The new measurements of the $b$ quark mass by ALEPH and DELPHI
are in good agreement with determinations at lower scales~\cite{Pich},
extrapolated to the Z pole by using the running predicted by 
perturbative QCD, as shown in figure \ref{fig1}. There are some indications 
for still uncontrolled biases from hadronization 
and/or uncomputed higher orders, since the results based on the three-jet
rate tend to be systematically lower than the one obtained from the first
moment of an event shape distribution.


\end{document}